\documentclass[%
 aip,
 pop
 amsmath,amssymb,
 reprint,%
]{revtex4-1}

\usepackage{graphicx}
\usepackage{dcolumn}
\usepackage{bm}

\newcommand{\sss}{\scriptscriptstyle}

\newcommand {\be}{\begin{equation}} 
\newcommand{\ee}{\end{equation}}    

\begin{document}
\preprint{AIP/123-QED}

\title[]{Energy  in density gradient  }

\author{J. Vranjes}
  \email{jvranjes@yahoo.com}
  \affiliation{
Instituto de Astrofisica de Canarias, 38205 La Laguna, Tenerife, Spain and \\
Departamento de Astrofisica, Universidad de La Laguna, 38205 La Laguna, Tenerife, Spain
}%
\author{M. Kono}%
 \email{kono@fps.chuo-u.ac.jp}
\affiliation{
Faculty of Policy Studies, Chuo University, Tokyo, Japan
}%
\date{\today}

\begin{abstract}
 Inhomogeneous plasmas and fluids contain energy stored in inhomogeneity and they naturally tend to relax into lower energy states by developing instabilities or by diffusion.  But the actual amount of energy in such inhomogeneities has remained unknown.  In the present work the amount of energy stored in a density gradient is calculated  for several specific density profiles in a cylindric configuration. This is of practical importance for drift wave instability in various plasmas, and in particular in its application in models dealing with the heating of solar corona because the  instability is accompanied with stochastic heating, so the energy contained  in inhomogeneity is effectively transformed into heat. It is shown that even for a rather moderate increase of the density at the axis in magnetic structures in the corona by a factor   1.5 or 3, the amount of excess energy per unit volume stored in such a density gradient becomes several orders of magnitude  greater than the amount of total energy losses per unit volume (per second)  in quiet regions in the corona. Consequently, within the life-time of a magnetic structure such  energy losses can  easily  be compensated  by the stochastic drift wave heating.

 \end{abstract}

\pacs{52.35.Kt; 52.50.-b; 96.60.-j; 96.60.P-   }
\maketitle



%

\section{Introduction }\label{s1}

From standard theory it is known that an isothermal system with the volume $V$, consisting of several species $j$ and with the total number of  particles for each species  $N_j$, contains a total amount of internal energy in the given volume $U_{int}\equiv U_{t}=(3 \kappa T/2)\sum_j N_j$, which is just the thermal energy of the system.

This internal  energy may be modified in various ways. For example, if the species are charged and they include electrons and ions, the chaotic thermal motion is affected by the Coulomb interaction. The internal energy in this case contains such a contribution and it is reduced:
\be
U_{int}=U_{t} \left[1-\frac{1}{12 \pi} \left(\frac{d}{r_d}\right)^3\right]. \label{e1}
\ee
Here, $d$ is the mean distance between the charged particles and $r_d$ is the plasma Debye radius. Such a modification of the internal energy implies also modifications of all other thermodynamic functions; i.e.,  the free energy, pressure, and entropy are all reduced, and the heat capacity increased. One example of this kind, dealing with dusty plasmas, may be found in  \citet{pv}. In most  of  electron-ion plasmas the Coulomb correction is small (implying a relatively large amount of particles within the Debye sphere), and such plasmas are then called ideal, or weakly non-ideal ones.

The total internal energy (and thus all other thermodynamic functions as well) may change also in the presence of some external forces. Without such forces a gas or plasma system  will tend to relax into an energetically most favorable state with minimum energy, and this usually means it will become isothermal and homogeneous. However, these external forces may cause gradients of various quantities, like density, temperature etc., and the internal energy (\ref{e1}) and other thermodynamic functions will change. Classic studies dealing with the free energy and thermodynamics of an inhomogeneous environment may be found in works \citet{ch, har, rc, sil, mc, war}, and \citet{lb}. These works mainly contain a general theory while for  practical purposes  some explicit results for the free energy are needed, and such an analysis will be performed in the present work.

The presence of such gradients implies an excess of energy, and the system will tend to get rid of it through  diffusion of particles until the minimum energy state is achieved, or much more efficiently by developing various instabilities, including drift instabilities as one example of general interest for plasmas.
 Drift instabilities in an inhomogeneous environment represent an efficient  way of relaxation towards a lower energy state, and they are present in almost every realistic plasma configuration. In the laboratory plasmas they represent a major issue  for confinement, though  the actual amount of energy stored in the plasma inhomogeneity has not been calculated so far.   For various purposes  it may be very useful and instructive  to know in advance  the total amount of this extra energy.
 For example, this may  help in modeling  stochastic heating based on such drift instabilities, in particular in the solar corona environment,
 suggested recently in a series of our works.\citep{v1, v2, v3, v3a, v4, v5}

 In the present work the excess energy stored in the density gradient is    calculated for some relatively simple  density profiles. The obtained results should be  applicable to various plasmas both in the laboratory and in space, and to fluids in general.


\section{Energy in density gradient }\label{s2}

We shall assume a cylindric volume $V=\pi R^2 L_z$ inhomogeneous due to  some external forces, where $R$ is the radius and $L_z$ is the axial  length. For example, in the presence of a magnetic field the radial density gradient, or pressure gradient in general, may be counteracted by the magnetic pressure. This in principle implies the presence of the radial inhomogeneity of the magnetic field as well, which can be small in case of a small plasma-beta, \citep{v11,v22} yet in any case it is supposed to balance the assumed pressure (density) gradient. But quite generally, the quasi-static equilibrium   may be described  through the  balance of forces
 \be
 0=-\nabla p+\sum_l \vec F_l. \label{e2}
 \ee
  So we may proceed by taking such a density gradient as a fact, knowing that it must imply additional forces acting on the plasma, and this means that it is  not in the minimum energy state. In other words, once such external forces are removed, or due to collisions and diffusion, the plasma will naturally  relax into the lowest possible energy state, occupying thus the whole volume $V$ and having some  constant number density $n_h$ and a constant temperature.
 \begin{figure}[!htb]
   \centering
  \includegraphics[height=6.5cm,bb=16 16 266 213,clip=]{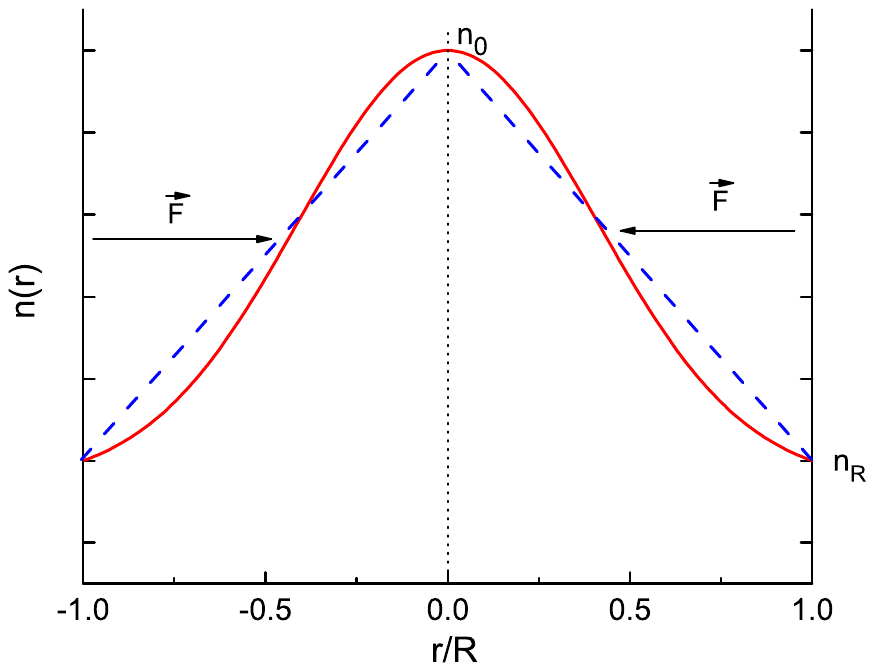}
      \caption{Sketch of the two density profiles, linear and Gaussian, in a cylindric volume $V=\pi R^2 L_z$, in balance with some radially acting force $\vec F$.  }   \label{f1}
       \end{figure}
Our inhomogeneous system thus has  some radially dependent  density $n(r)$, and we shall calculate its  internal energy (which includes thermal and potential energy)  taking several possible cases for the density shape.

\begin{figure}[!htb]
   \centering
  \includegraphics[height=6cm,bb=66 400 530 774,clip=]{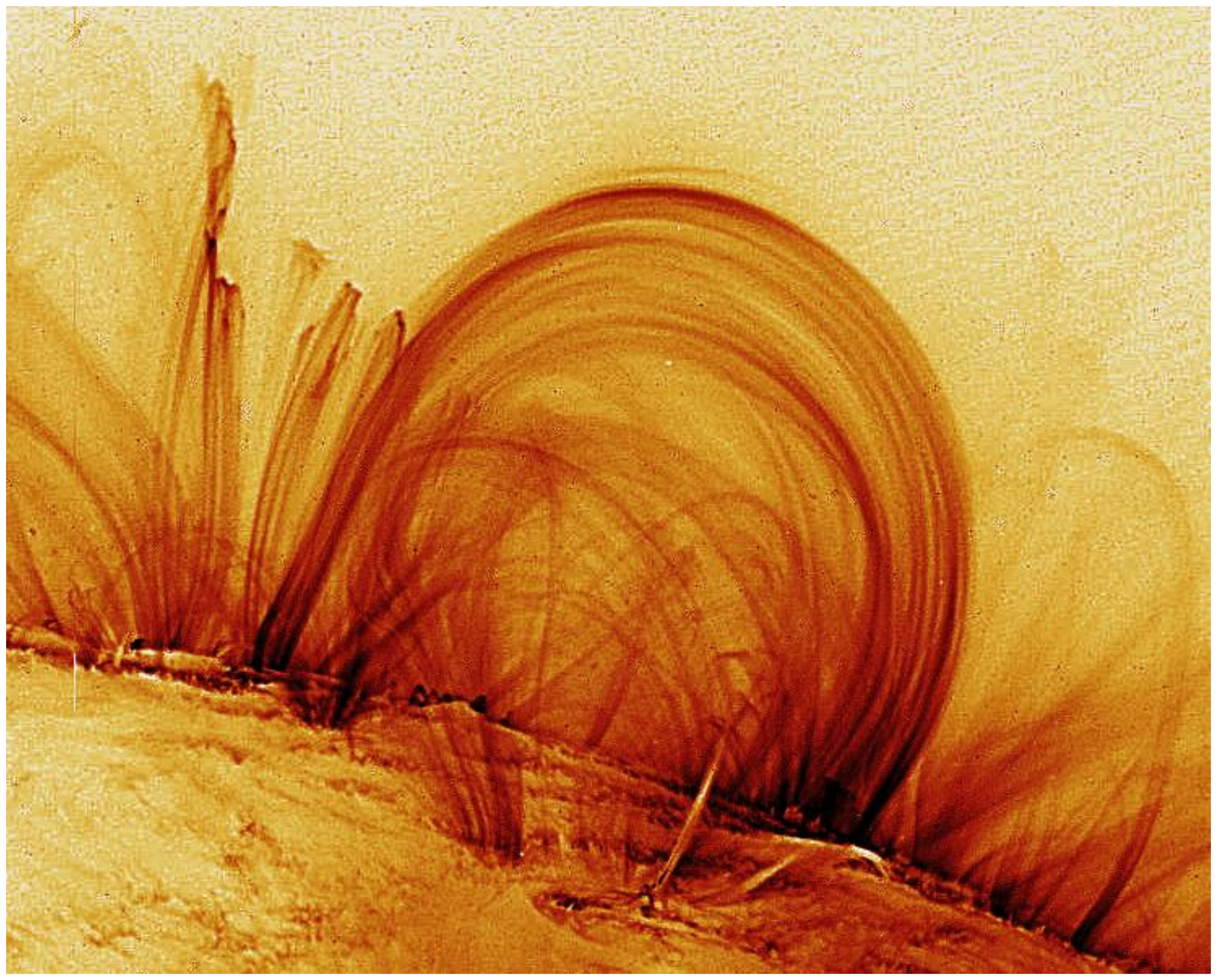} \caption{Coronal magnetic structures, picture  taken by the Transition Region and Coronal Explorer (TRACE).\citep{trace}} \label{f22}
\end{figure}

\begin{figure}[!htb]
   \centering
  \includegraphics[height=5.7cm,bb=68 560 353 774,clip=]{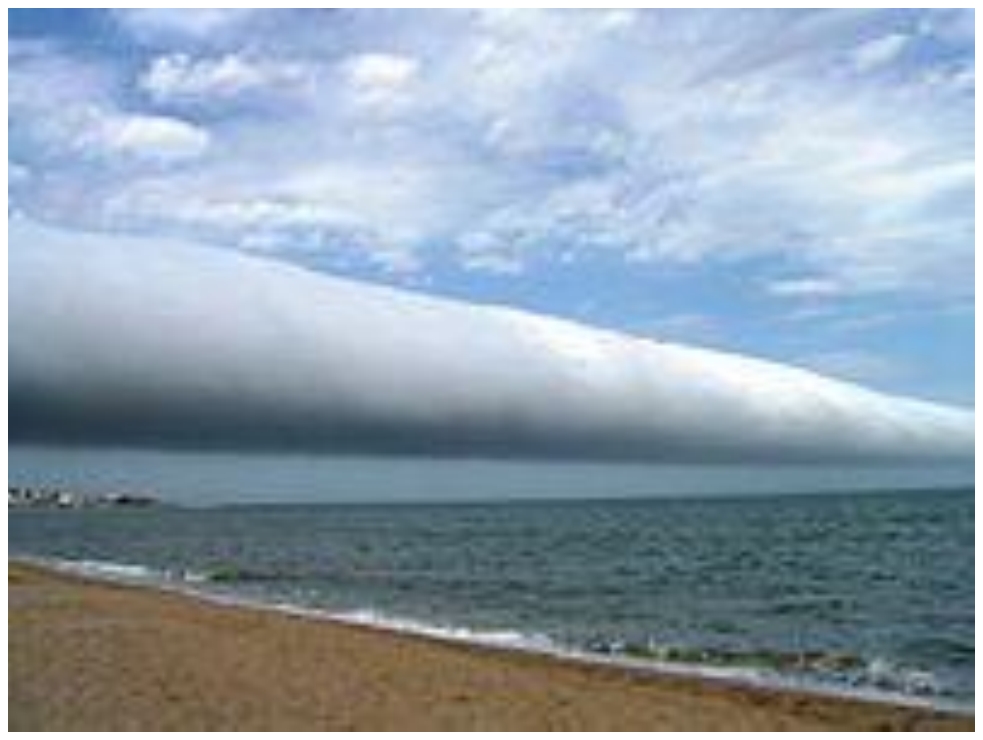} \caption{A roll cloud pictured in Uruguay. \citep{wiki}} \label{f33}
\end{figure}

\begin{figure}[!htb]
   \centering
  \includegraphics[height=5.6cm,bb=68 666 225 774,clip=]{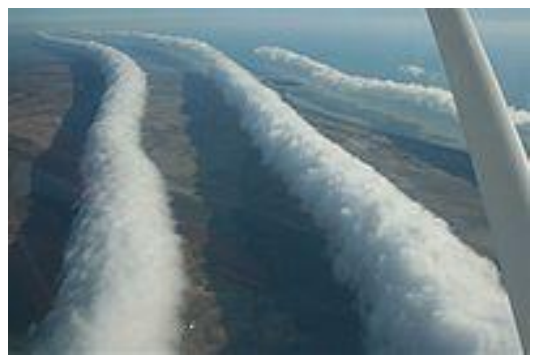} \caption{Multiple roll clouds pictured in Australia. \citep{wiki2}} \label{f44}
\end{figure}

The theory presented here may  be applicable to coronal magnetic structures\citep{trace} [see Fig.~\ref{f22}], and to some atmospheric phenomena like roll clouds and bores. Roll clouds  are   tube-shaped,  many kilometers long structures rolled up about a horizontal axis, single \citep{wiki} or multiple \cite{wiki2} [see Figs.~\ref{f33},~\ref{f44}], like so called  Morning Glory cloud frequently observed in  Australia\citep{smi} that can be between hundred and thousand kilometers long and only about one  kilometer in diameter,\citep{grim, gol} and also clouds observed in South America and elsewhere.\citep{wiki, hart}  A clear pressure jump of up to 200 Pa in such rolling solitons has been measured, \citep{gol} implying the excess of energy in them as compared with the surrounding fluid. Yet another atmospheric example of similar kind are tornados.

In derivations  further in the text, the potential and internal energies are per species, so the theory is valid for ordinary fluids, and in electron-ion plasmas (or multi-component in general with a number of species $\mu$) all expressions should be multiplied by a factor 2 (or by factor $\mu$).

\subsection{Linear density profile with porous boundaries}\label{ss1}

We shall assume a nearly ideal plasma with a negligible Coulomb interaction energy and having  a  linear number density profile
 \be
 n(r)=n_0 - a r. \label{lin}
 \ee
  The number density has the value $n_{\sss R}$ at the edge of the cylinder with the radius $R$, and it increases  towards the axis to a maximum value $n_0$, see Fig.~\ref{f1}, the dashed line.  This means that $a=(n_0-n_{\sss R})/R$.  It will be shown later that  the expected results for the internal energy should not be much different from some more realistic cases, like the Gaussian profile presented in the same figure. Note that in some cases, like in some  magnetic configuration in the solar plasma, the profile may be just inverted, with a decreasing density towards the axis, implying forces acting in the opposite direction. However, there will again be a  resulting potential energy stored in the inhomogeneity.

If the forces are removed, or due to collisions, in time the system will relax to the minimum energy homogeneous state with some  constant density $n_h$. In space or in solar plasma we may assume that this  extra density in the gradient from Fig.~\ref{f1}  will be absorbed by the infinite reservoir of the plasma around with the assumed density  $n_{\sss R}$. So the potential energy will be calculated with respect to $n_{\sss R}$, and this we shall call the model of  {\em  porous or absorbing boundaries}. In such environments this is rather justified.  However, in bounded plasmas and fluids in the laboratory, the relaxed density will clearly have some value  $n_h>n_{\sss R}$. This issue  shall be addressed later completely exactly in Sec.~\ref{na}.

      The total number of particles in the volume $V$ is $N=\int n(r) dV$, where $dV= r dr d\theta dz$ and integration is over the angle $\theta\in (0, 2\pi)$, over $z\in (0, L_z)$, and $r\in (0, R)$. Eventual variation of density in $z$ and $\theta$ directions is neglected. The result is:
  \be
  N=V\frac{n_0 + 2 n_{\sss R}}{3}. \label{e3}
  \ee

  The thermal energy for the case of inhomogeneous density is $U_t=\int (3 \kappa T/2) n(r) dV$. For the density (\ref{lin}) and with the boundary conditions $n(0)=n_0$, $n(R)=n_{\sss R}$  this yields
  \be
U_t=V \kappa T\left(\frac{n_0}{2}+ n_{\sss R}\right). \label{e5}
\ee
 In  view of Eq.~(\ref{e2}), the internal energy in the present case must be the sum of the thermal energy $U_t$ and some potential energy which system has achieved due to the work of the external forces $\sum_l \vec F_l$. In principle, presently we are not interested in separate forces $\vec F_l$,  but we shall use the fact that their total sum  exactly matches the pressure force. So in order to calculate the potential energy it may be convenient to use the pressure force instead of $\sum_l \vec F_l$, and express the potential energy through the temperature and density only.

 The potential energy $E_1(r)$ of a unit volume at some position $r$, obtained by the act of forces acting on it, is the integral over the distance and it can be written as
   \be
    E_1(r)=\int^r\sum_l\vec F_l(r') d\vec {r'}=\int^r \!\!\nabla p d\vec {r'}=\kappa T n_1(r). \label{e3a}
    \ee
The number density $n(r)$  has the lowest value $n_{\sss R}$ at the boundary, and in case of some large  configurations in the atmosphere or in space, like solar magnetic structures this value will be assumed the same  outside for $r>R$ (porous or absorbing boundary). Inside the structure, clearly  the excess  density above $n_{\sss R}$ will contain  additional potential energy due to action of forces, see Fig.~\ref{f1}.  Thus the density $n_1$ in  Eq.~(\ref{e3a})  satisfies  the following boundary conditions: $n_1(0)=n_0- n_{\sss R}$, and $n_1(R)=0$. This yields $n_1(r)=(n_0-n_{\sss R}) (1- r/R)$.

The total potential energy in the volume $V$ is thus the integral of $E_1(r)$ over the whole volume and it reads:
\be
E_p=V \kappa T\frac{n_0-n_{\sss R}}{3}.
\label{e4}
\ee

The total internal energy is the sum of the two, $U_{int}=U_t+ E_p$, and the Gibbs-Helmholtz thermodynamic free energy $F$ can easily be obtained using the well known relation
 \[
  U_{int}=-T^2\frac{\partial}{\partial T}\left(\frac{F}{T}\right)_{V}.
  \]
    The specific heat and entropy are given by $C_v=\partial U_{int}/\partial T|_{\sss V}$, $S=-\partial F/\partial T|_{\sss V}$. However, for  the drift instabilities only the potential part of the internal energy may play the role, and Eq.~(\ref{e4}) shows that it is present as long as $n_{\sss R}\neq n_0$.

\subsection{Gaussian and arbitrary flat-top  density profile }\label{ss2}

For a Gaussian density profile (the full line in Fig.~\ref{f1}) $n(r)=c_1 \exp(- a r^2) + c_2$, where $c_1, c_2$ are chosen to satisfy the boundary conditions
$n(0)=n_0$, $n(R)=n_{\sss R}$, we have
\be
n(r)=n_0- b \left[1-\exp(-a r^2)\right], \quad b=\frac{n_0-n_{\sss R}}{ 1-\exp(-a R^2)}. \label{g1}
\ee
This is used to calculate the thermal energy. The result is
\be
U_t=\frac{3}{2}V \kappa T [\alpha (n_0-n_{\sss R}) + n_{\sss R}], \label{ut}
\ee
\[
\alpha=1+ \frac{1}{a R^2} - \frac{1}{1- \exp(-a R^2)}.
\]
For the potential energy the boundary conditions are as before $n_1(0)=n_0- n_{\sss R}$, $n_1(R)=0$, i.e., the density (\ref{g1}) is reduced by  $n_{\sss R}$. The mean potential energy per unit volume now becomes
\be
\frac{E_p}{V}=\kappa T \alpha (n_0-n_{\sss R}). \label{g2}
\ee
Comparing this with Eq.~(\ref{e4}) it is seen that the potential energy in the present case may be greater than the value
obtained for the linear density profile providing that $\alpha>1/3$.
Note that $\alpha\rightarrow 0.5$ when $aR^2\rightarrow 0$, and $\alpha\rightarrow 0$ when $aR^2\rightarrow \infty$, and $\alpha\approx 0.42$ when $aR^2\approx 1$. So the differences between the two profiles are not drastic in any case.
The total internal energy is consequently
\[
 U_{int}=U_t+ E_p=\frac{V \kappa T}{2}\left[5 \alpha (n_0-n_{\sss R})+ 3 n_{\sss R}\right].
 \]
Instead of the Gaussian density we may have  an arbitrary flat-top profile which can be found in numerous plasmas $n(r)=c_1 \exp(- a r^k) + c_2$, where again $c_1, c_2$ are chosen to satisfy $n(0)=n_0$, $n(R)=n_{\sss R}$. For $k=2$ this yields the Gaussian case discussed above, and in the limit $k\rightarrow \infty$ it  gives a step-profile.
Repeating the calculation presented above we obtain the potential energy density (\ref{g2}) where now
\[
\alpha=1-\frac{1}{1-\exp(- a R^k)}
\]
\be
 -\frac{2}{R^2}\frac{1}{1-\exp(- a R^k)} \frac{a^{-2/k}}{k} \left( \Gamma[2/k, a r^k]\right)^R_0, \label{g22}
\ee
and $\Gamma[b, x]$ is the  incomplete Euler gamma  function.

In fact, the analysis can easily be generalized to any density profile in the cylindric system, bearing in mind the procedure described above. The result for the potential energy is
\be
E_p=V \frac{2 \kappa T}{R^2}\int_0^R\left[n(r)- n_{\sss R}\right] r dr, \label{g4}
\ee
 and the thermal energy is $U_t=3 \pi \kappa TL_z\int_0^Rn(r)r dr=(V 3\kappa T/R^2)\int_0^R n(r) r dr$.

\subsection{Application to heating in the solar corona }\label{ss3}
The solar corona is hot and it continuously looses its energy. Without an efficient generator of energy, the amount expressed through
Eq.~(\ref{e1}) would be lost in less than a day and the corona would become completely cool. However, it remains hot and with the temperature
exceeding a million degrees.

The presence of the density gradient implies drift instabilities and those may lead to stochastic heating if certain conditions are satisfied. \citep{hat, mc1, mc3, mc2}  We shall  make some estimate and application of the present  results to the drift wave heating paradigm in the solar atmosphere,  which we put forward recently.\citep{v1, v2, v3, v3a, v4, v5}

For {\em the quiet regions} in the solar corona, presently accepted required heating rate, \citep{nar} with  the temperature  $T=1.8\cdot 10^6$ K, is  in the range $\Gamma_0\equiv E_0/s=2\cdot10^{-7}$ J/(m$^3$s) to $\Gamma_0=4\cdot10^{-6}$ J/(m$^3$s) for the two number densities  $n_q=0.2\cdot 10^{15}$ m$^{-3}$ and $n_q= 10^{15}$ m$^{-3}$, respectively. Note that for such parameters the corrections due to Coulomb interaction in Eq.~(\ref{e1}) are completely negligible.
Such a  heating rate can easily be satisfied in the solar magnetic structures following the models based on the drift wave stochastic heating.\citep{v1, v2, v3, v3a, v4, v5} In these studies  various heating rate values are  obtained, dependent on the plasma parameters. As example, in magnetic structures with the characteristic inhomogeneity scale-length $L_n$ of around several hundred kilometers, the wave frequency is of the order of 0.1 Hz and the energy release rate $\Gamma$ due to the stochastic heating  is around the value $\Gamma_0$ given above, see more in \citet{v1}.

   We may now take some specific   density profile and calculate the potential energy density $E_1=E_p/V$  for the heating rate $\Gamma\simeq \Gamma_0$  in order to see at least roughly for how long such a density profile might sustain the plasma at the given temperature, i.e., for how long it may  compensate for the energy losses. The result is presented in Table~\ref{t1} and Table~\ref{t2} for the temperature and two starting densities for quiet regions $n_{\sss R}=n_q$ given above, and allowing for several possible density values $n_0$ in the center of the loop. From Table~\ref{t1} it is seen that even in the case when the boundary value $n_{\sss R}$ increases only by a factor 1.5 at the axis,  the mean potential energy density becomes  $E_1=E_p/V=8.3\cdot 10^{-4}$ J/m$^3$, which is several orders of magnitude greater than the energy lost per unit volume in one second. This is even more so for the values in Table~\ref{t2}.    But in reality, the density gradient changes (i.e., the density profile flattens) in time and so does the realistic energy release rate; observations show that the mean life time of the loops is in the range of half an hour to a few hours. So to  make some estimate we may set the loop life-time  equal one hour, and to have a realistic margin and to be sure about a sustainable heating it would be  necessary to have the ratio $E_1/E_0$  considerably greater than 3600. The numbers given  in Tables~\ref{t1}, \ref{t2} show that this is indeed the case for almost all values of $n_0$.
Observe also that the potential energy becomes a considerable part of the total internal energy already for  $n_{\sss R}/n_0>3$. As stressed before, the numbers presented here are per species.

   A similar analysis can be performed for the coronal holes where $\Gamma_0=8\cdot10^{-6}$ J/(m$^3$s), showing that the amount of energy in the density gradient is more than enough for a sustainable heating.
 \begin{table}[]
\caption{\label{t1}Parameters for  quiet regions following \citet{nar} with $n_{\sss R}=0.2\cdot 10^{15}$ m$^{-3}$ and for linear density profile given by Eq.~(\ref{lin}),  with several values $n_0$ at the center.     }
\begin{ruledtabular}
\begin{tabular}{lcccc}
$n_0/n_{\sss R}$ & 1.5 & 3 & 5 & 10\\
$E_1$ [J/m$^3$]& 0.0008 & 0.0033 & 0.0066 & 0.015\\
$U_{th}/V$ [J/m$^3$]&0.0087 & 0.012& 0.017 & 0.03\\
$E_p/U_{th}$&0.095 & 0.27& 0.38 & 0.5\\
$E_1/E_0$& 4142 & 16568 & 33137 & 74558\\
\end{tabular}
\end{ruledtabular}
\end{table}

 \begin{table}[]
\caption{\label{t2}Parameters for  quiet regions following \citet{nar} with $n_{\sss R}=10^{15}$ m$^{-3}$ and for linear density profile (\ref{lin}). }
\begin{ruledtabular}
\begin{tabular}{lcccc}
$n_0/n_{\sss R}$ & 1.5 & 3 & 5 & 10\\
$E_1$ [J/m$^3$]& 0.0041 & 0.0165 & 0.033 & 0.074\\
$U_{th}/V$ [J/m$^3$]&0.043 & 0.062& 0.087 & 0.15\\
$E_p/U_{th}$&0.095 & 0.27& 0.38 & 0.5\\
$E_1/E_0$& 20710 & 82842 & 165684 & 372789\\
\end{tabular}
\end{ruledtabular}
\end{table}
 In {\em active regions}, following \citet{nar} the temperature and the number density vary from  $T=10^4$ K, $n_a=0.5\cdot 10^{15}$ m$^{-3}$     to
 $T=2.5 \cdot 10^6$ K, $n_a=5\cdot 10^{15}$ m$^{-3}=n_{\sss R}$, and the energy losses are in the range $\Gamma_0=7\cdot10^{-5}$ J/(m$^3$s) to $\Gamma_0=3\cdot10^{-4}$ J/(m$^3$s). For the  first set of data a sustainable heating  $\Gamma\simeq \Gamma_0$ appears possible for a few second only, provided that $n_0=10 n_{\sss R}$. However, for the second set $T=2.5 \cdot 10^6$ K, $n_a=5\cdot 10^{15}$ m$^{-3}$ the situation is a bit different, and the result is presented in Table~\ref{t3}. These data indicate that a rather strong density gradient is needed to  sustain the energy losses within the life-time of a magnetic structure in active regions.

 \begin{table}[]
\caption{\label{t3}Parameters for  active  regions following \citet{nar} with $T=2.5 \cdot 10^6$ K, $n_a=5\cdot 10^{15}$ m$^{-3}$,  for linear density profile (\ref{lin}), and with several values $n_0$ at the center.   }
\begin{ruledtabular}
\begin{tabular}{lccccc}
$n_0/n_{\sss R}$ & 1.5 & 3 & 5 & 10 & 15\\
$E_1=E_p/V$ [J/m$^3$]& 0.03 & 0.115 & 0.23 & 0.52& 0.18\\
$U_{th}/V$ [J/m$^3$]&0.3 & 0.43& 0.6 & 1.04& 1.47\\
$E_p/U_{th}$&0.095 & 0.27& 0.38 & 0.5& 0.55\\
$E_1/E_0$& 96 & 383 & 767 & 1726& 2685\\
\end{tabular}
\end{ruledtabular}
\end{table}

 The analysis presented above can be repeated for the Gaussian density profile (\ref{g1}). Without going into details, we may take the case $aR^2\approx 1$,  and  in view of Eqs.~(\ref{e4}, \ref{g2}) it may be shown that the above given values for the heating time should be multiplied by a factor 1.26. So the heating in this case in quiet regions and coronal holes is even more certain and this holds for all cases with $aR^2< 1$.

 The calculated potential energy stored in the density gradient and the consequent stochastic heating by the drift wave are thus more than enough to compensate for losses in quiet Sun regions and in coronal holes, and the predicted  sustainable stochastic heating\citep{v1, v2, v3, v3a, v4, v5} by the drift wave looks like a rather realistic scenario. In active regions it may sustain heating only provided relatively strong density gradients as Table~\ref{t3} suggests, so most likely some additional and more energetic  processes are in action in such regions.

\section{Non-absorbing boundaries }\label{na}

\begin{figure}[!htb]
   \centering
  \includegraphics[height=6.5cm,bb=16 14 267 207,clip=]{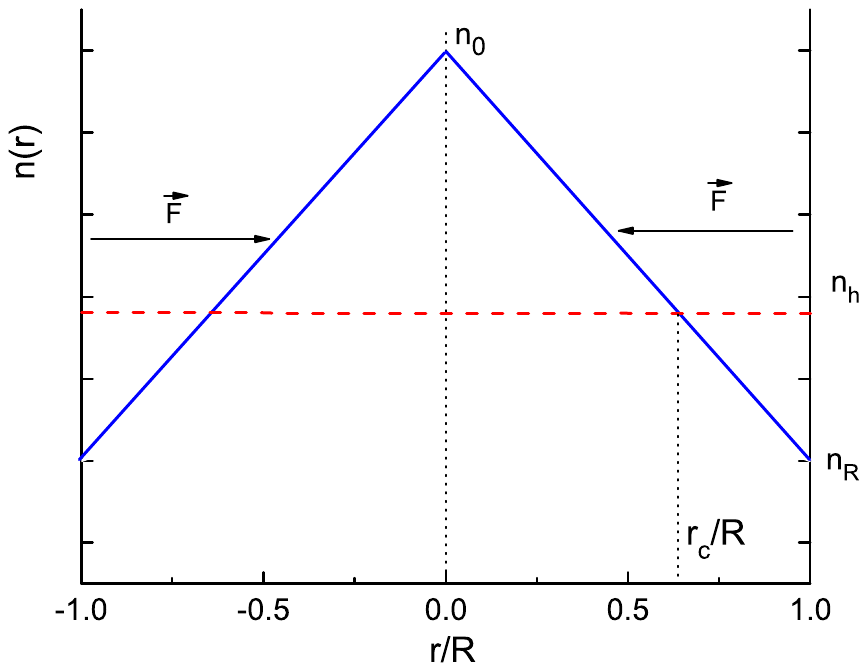}
      \caption{Sketch of the density profile in a cylindric volume $V=\pi R^2 L_z$ for a non-absorbing boundary. Without forces the system relaxes to a homogeneous state with $n=n_h$ where $n_{\sss R}< n_h< n_0$.   }   \label{f2}
       \end{figure}

\begin{figure}[!htb]
   \centering
  \includegraphics[height=6.5cm,bb=14 13 249 197,clip=]{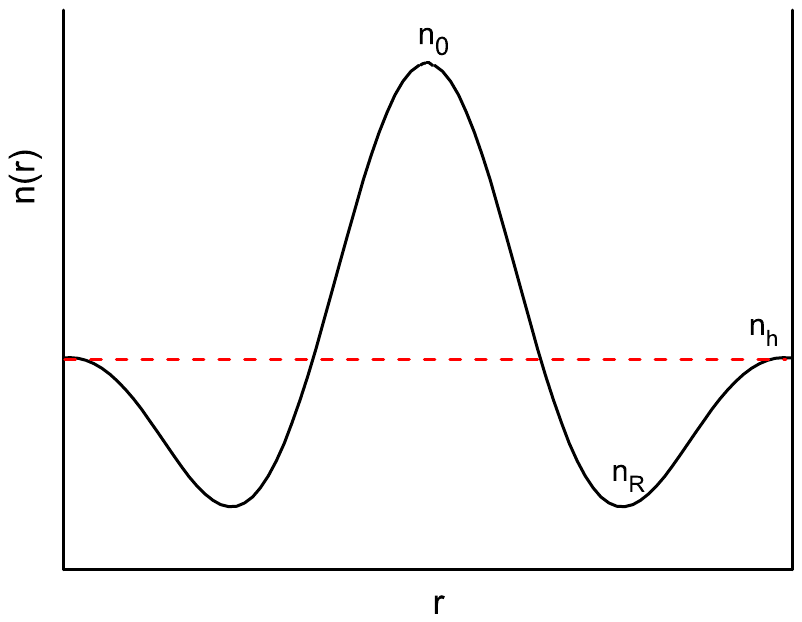}
      \caption{Sketch of the two density profiles, linear and Gaussian, in a cylindric volume $V=\pi R^2 L_z$, in balance with some radially acting force $\vec F$. Without forces the system relaxes to a homogeneous state with $n=n_h$.   }   \label{f2b}
       \end{figure}

In the laboratory environment  the initial density profile with the  number density  $n_{\sss R}$ at the edge of the cylinder
will normally relax  to a homogeneous  minimum available energy  state with the density line $n=n_h$, see Fig.~\ref{f2}. We shall use this simple linear density for the present analysis although obviously it can easily be replaced with the Gaussian one as shown in Sec.~\ref{s2}. Quantitatively, the results will be similar. Clearly there must be $n_{\sss R}< n_h< n_0$  as depicted in Fig.~\ref{f2}.  The same situation may sometimes happen also in coronal magnetic structures with exceptionally strong boundary magnetic field when  diffusion across the boundary itself is much weaker as compared with the diffusion inside the structure and  when there are drift instabilities taking place inside the structure. In this case the density $n_h$ would correspond to the  density of the bulk plasma around the structure.  
A more realistic situation in coronal structures is  presented in Fig.~\ref{f2b} where we have the  external density $n_h$, and the given density inside the structure,  which is a more physical variant  of the rough profile from Fig.~\ref{f2}.

 Note that the assumed actual linear density profile in Fig.~\ref{f2} is in fact partly inverted with respect to $n_h$, and both domains $n(r)> n_h$ and $n(r)<n_h$ have some potential energy with respect to the relaxed state with $n=n_h$.

 The density $n_h$ can be calculated in the following manner. The total number of particles in the volume $V$ is again given by Eq.~(\ref{e3}).
  In the relaxed state, both the total number of particles and volume remain the same, so $N=n_h V$ and this directly yields
  \[
  n_h=\frac{n_0}{3}+ \frac{2 n_{\sss R}}{3}.
  \]
   The intersection of the lines $n=n(r)$ and $n=n_h$ is at $r_c= 2 R/3$ and this remains so for any $n_0$ and $n_{\sss R}$. At $t\rightarrow \infty$ the internal energy will reduce  to  the thermal energy (\ref{e5}). In any other moment it will be the sum of the internal energy (\ref{e5}) and potential energy $\Sigma_p$ with respect to the density level $n_h$. The potential energy is the sum of the two  integrals $\Sigma_p=\Sigma_{p1} + \Sigma_{p2}$,
    \[
    \Sigma_{p1}= \kappa T\int_0^{r_c} [n(r)-n_h] dV, \quad \Sigma_{p2}=\kappa T\int_{r_c}^R [n_h-n(r)] dV.
    \]
    The result turns out to be
    \[
    \Sigma_{p1}=\Sigma_{p2}=\frac{8}{81} V\kappa T (n_0- n_{\sss R}),
    \]
    so that
    \be
    \Sigma_p=\frac{16}{81} V\kappa T (n_0- n_{\sss R}). \label{b2}
    \ee
    This may be compared with the potential energy (\ref{e4}) from the case of absorbing boundaries. The ratio of the two is $E_p/\Sigma_p \approx 1.7$. Such a result could have been foreseen because the reference level with respect to which the energy is  calculated in the present case is higher, so the potential energy is naturally smaller.

   It should be stressed that this is a completely general and exact analysis  and it can     be performed in the same manner for any other analytically given density profile. For example, the profile in Fig.~\ref{f2b} is in fact a function of the kind $c_1+ \exp(- c_2 r^2) J_0(c_3 r)$ (where $J_0$ is the Bessel function), but such derivations will not be repeated here.

\section{Summary and conclusions }\label{s3}

Inhomogeneity of plasmas implies a free energy for development of drift instabilities as the most efficient way of relaxing
into a more favorable lower energy state. Such instabilities may be accompanied with stochastic heating provided large enough amplitudes
of the perturbations, and this phenomenon has been observed in the past in specially designed laboratory experiments.\citep{hat, mc1, mc3,  mc2}

The same idea has been used recently\citep{v1, v2, v3, v3a, v4, v5}  as a new paradigm for the heating of the solar corona. Analytical calculations in these references show
that such a stochastic heating can easily satisfy numerous heating requirements  like the necessary heating rate,  better heating of heavier particles, stronger heating in perpendicular direction with respect to the magnetic field vector, etc. The model is based on a quasi-static inhomogeneous background plasma which is supposed to contain enough energy in order for the mechanism to work. This assumption is checked in the present study, and the energy contained in the density gradient is calculated. The results obtained in the work suggest that  accidentally and externally created magnetic structures indeed contain enough  energy  for a sustainable heating at least in the solar quiet regions and in coronal holes.

 The inhomogeneity is an intrinsic feature of the magnetic structures in the solar atmosphere.  Movement and restructuring is  a continuous process in coronal magnetic structures and such phenomena are accompanied with the motion of plasma due to its frozen-in properties. The life-time of these structures is of the order of an  hour, so that indeed they can be treated as quasi-static for relatively fast drift instabilities, and they are  maintained by processes deep below the corona itself. This implies that such a  drift-wave-favorable environment is being created continuously;  structures appear and disappear all the time, and heating is taking place in each of them independently. As discussed in \citet{v1, v2}, the drift waves  and heating are produced directly in these magnetic structures and driven by the plasma inhomogeneity. However,  the complete  mechanism is totally  magnetic by nature:  such inhomogeneities are caused by the  dynamics and restructuring of the magnetic field, and these phenomena on the other hand are driven by processes far from the corona, i.e.,  inside the Sun.  In the present work we were focused only on the heating rate close to the energy loss rate, to check if it can be sustainable or not. However various heating rates are possible dependent on the drift wave frequency and growth-rate (which on the other hand are dependent on the inhomogeneity scale-length), including energy releases in the range of nano-flares\citep{v3}.

The analysis presented in the work is for a simple cylindric geometry, but it can easily be generalized
 to other  geometries applicable to the laboratory plasma like a tokamak,  or to loop structures in the solar atmosphere.

\vfill
\eject

\nocite{*}
\bibliography{aipsamp}

\begin{thebibliography}{99}

\bibitem[\protect\citeauthoryear{Pandey \&  Vranjes}{2005}]{pv} B. P. Pandey and J. Vranjes, Phys. Scripta {\bf 72}, 247 (2005).
\bibitem[\protect\citeauthoryear{Cahn \&  Hilliard}{1958}]{ch} J. W. Chan and J. E. Hilliard, J. Chem. Phys.  {\bf 28}, 258 (1958).

\bibitem[\protect\citeauthoryear{Hart}{1959}]{har} E. W. Hart, Phys. Rev. {\bf 113}, 412 (1959).

\bibitem[\protect\citeauthoryear{Rice \&  Chang}{1974}]{rc} O. K. Rice and D. R. Chang, Physica {\bf 78}, 500 (1974).

\bibitem[\protect\citeauthoryear{Silver}{1978}]{sil} R. N. Silver, Phys. Rev. B {\bf 17}, 3955 (1978).

\bibitem[\protect\citeauthoryear{McCoy \&  Davis}{1979}]{mc} B. F. McCoy and H. T. Davis, Phys. Rev. A {\bf 20}, 1201 (1979).

\bibitem[\protect\citeauthoryear{Warner}{1980}]{war} M.  Warner, Chem. Phys. Lett. {\bf 70}, 155 (1980).

\bibitem[\protect\citeauthoryear{Lowett \&  Baus}{1992}]{lb} R. Lowett and M. Baus, Physica A {\bf 181}, 309 (1992).


\bibitem[\protect\citeauthoryear{Vranjes \&  Poedts}{2009a}]{v1} J. Vranjes  and S. Poedts,  MNRAS  {\bf 398}, 918 (2009a).
\bibitem[\protect\citeauthoryear{Vranjes \&  Poedts}{2009b}]{v2} J. Vranjes  and S. Poedts,  MNRAS {\bf  400}, 2147 (2009b).
\bibitem[\protect\citeauthoryear{Vranjes \&  Poedts}{2009c}]{v3} J. Vranjes  and S. Poedts, Phys. Plasmas 16, 092902 (2009c).
\bibitem[\protect\citeauthoryear{Vranjes \&  Poedts}{2009d}]{v3a} J. Vranjes  and S. Poedts, EPL {\bf 86}, 39001  (2009d).
\bibitem[\protect\citeauthoryear{Vranjes \&  Poedts}{2010a}]{v4} J. Vranjes  and S. Poedts,  MNRAS {\bf 408}, 1835 (2010a).
\bibitem[\protect\citeauthoryear{Vranjes \&  Poedts}{2010b}]{v5} J. Vranjes and S. Poedts,  Astrophys. J. {\bf 719}, 1335 (2010b).


\bibitem[\protect\citeauthoryear{Vranjes}{2011}]{v11} J. Vranjes,    MNRAS {\bf 415}, 1543 (2011).
\bibitem[\protect\citeauthoryear{Vranjes}{2011}]{v22} J. Vranjes,  A\&A {\bf 532}, A137 (2011).

\bibitem[\protect\citeauthoryear{TRACE}{2014}]{trace} http://soi.stanford.edu/results/SolPhys200/Schrijver\-/TRACEpod\_archive.html

\bibitem[\protect\citeauthoryear{Wikipedia}{2014}]{wiki}  Wikipedia contributors. ``Arcus cloud." Wikipedia, The Free Encyclopedia. Wikipedia, The Free Encyclopedia, 1 Jun. 2014. Web. 19 Sep. 2014.

\bibitem[\protect\citeauthoryear{Wikipedia}{2014}]{wiki2} Wikipedia contributors. ``Morning Glory cloud." Wikipedia, The Free Encyclopedia. Wikipedia, The Free Encyclopedia, 30 Aug. 2014. Web. 19 Sep. 2014.

\bibitem[\protect\citeauthoryear{Smith}{1988}]{smi} R. K. Smith, Earth Sci. Rev. {\bf 25}, 267 (1988).


\bibitem[\protect\citeauthoryear{Rottman \& Grimshaw}{2004}]{grim} J. W. Rottman and R. Grimshaw, in {\em Environmental stratified flows}, ed. R. Grimshaw (Kluver, Dordrecht, 2003).

\bibitem[\protect\citeauthoryear{Goler \& Reeder}{2004}]{gol} R. A.  Goler and M. J. Reeder, J. Atmos. Sci. {\bf 61}, 1360 (2004).

\bibitem[\protect\citeauthoryear{Hartung \& Sitkowski}{2010}]{hart} D. C. Hartung and M. Sitkowski, Weather {\bf 65}, 148 (2010).


\bibitem[\protect\citeauthoryear{Hatakeyama et al.}{1980}]{hat} R. Hatakeyama, M. Oertl, E. M\"{a}rk, and R. Schrittwieser, Phys. Fluids {\bf  23},
1774 (1980).
\bibitem[\protect\citeauthoryear{McChesney et al.}{1987}]{mc1}  J. M. McChesney,  R. A. Stern, and P. M.  Bellan,  Phys. Rev. Lett. {\bf 59},
1436 (1987).
\bibitem[\protect\citeauthoryear{McChesney et al.}{1991}]{mc3} J. M. McChesney, P. M. Bellan, and R. A.  Stern,  Phys. Fluids B {\bf 3}, 3363 (1991).
\bibitem[\protect\citeauthoryear{Sanders et al.}{1998}]{mc2}  S. J. Sanders, P. M. Bellan, and  R. A. Stern,  Phys. Plasmas {\bf 5}, 716 (1998).

\bibitem[\protect\citeauthoryear{Narain \&  Ulmschenider}{1990}]{nar} U. Narain and P. Ulmschneider, Space. Sci. Rev.  {\bf 54}, 377 (1990).




\end{thebibliography}

\end{document}